\documentstyle[aps,prl,psfig,twocolumn]{revtex}

\def\be{\begin{equation}}
\def\ee{\end{equation}}
\def\bea{\begin{eqnarray}}
\def\eea{\end{eqnarray}}
\def\bma{\begin{mathletters}}
\def\ema{\end{mathletters}}

\def\C{\hbox{$\mit I$\kern-.7em$\mit C$}}
\def\st{\mbox{ s. t. }}
\newcommand{\one}{\mbox{$1 \hspace{-1.0mm}  {\bf l}$}}
\def\bg{\begin{guess}}
\def\eg{\end{guess}}
\newcommand{\bra}[1]{\mbox{$\langle #1 |$}}
\newcommand{\ket}[1]{\mbox{$| #1 \rangle$}}
\newcommand{\kepsi}{\mbox{$|\psi^{< k} \rangle$}}
\newcommand{\brapsi}{\mbox{$\langle \psi^{< k}|$}}
\newcommand{\bracket}[2]{\mbox{$\langle {{#1}} \mathrel{ | {\vphantom
        {{#1} {#2}}} \kern-\nulldelimiterspace} {{#2}} \rangle$}}

\newcommand{\rem}[1]{}

\newcommand{\calp}{\mbox{$\cal P$}}

\begin{document}
\draft

\title{Schmidt number witnesses and bound entanglement}

\author{Anna Sanpera, Dagmar Bru\ss\ and Maciej Lewenstein}

\address{
Institute f\"ur Theoretische Physik,
Universit\"at Hannover, 30169 Hannover,Germany}

\date{\today}

\maketitle

\begin{abstract}

The Schmidt number of a mixed state characterizes the minimum Schmidt
rank of the pure states needed to construct it. 
We investigate the Schmidt number of  an arbitrary mixed state by 
constructing a Schmidt number witness that detects it.
We present a canonical form of such witnesses and
provide constructive methods for their
optimization.  Finally, we present  strong
evidence that all bound entangled states with positive partial transpose in
${\cal H}_3\otimes {\cal H}_3$ have Schmidt number 2.
\end{abstract}


\narrowtext

Characterization of entanglement is one of the key features related to quantum 
information theory\cite{rew2000}. 
The resources needed to  implement a particular protocol of quantum information processing
(e.g. \cite{ekert})
are closely linked to the entanglement properties of the states used 
in the protocol. 
Although recently a great effort has been devoted to detect the 
presence of entanglement in a given state (see for instance 
\cite{review,lewen002}) 
and also to characterize multipartite entangled systems\cite{multien}, many 
questions concerning bipartite mixed systems remain unanswered.

A bipartite pure state
$\ket{\psi}$ can always be described by its
Schmidt decomposition; i.e. the representation of $\ket{\psi}$ 
in an orthogonal product basis with minimal number of terms. 
The Schmidt rank is the number
of non-vanishing terms in such an expansion. 
This decomposition gives a clear insight
on  the number of degrees of freedom that are entangled
between both parties, and its coefficients provide a
measure of entanglement.

The characterization of mixed states is a much 
harder task, and despite the fact that many entanglement measures 
have been introduced \cite{measures}, there is not a ``canonical'' way of 
quantifying the entanglement. Nevertheless, in the context of mixed 
bipartite states it is legitimate and meaningful 
to ask:  which is the minimum number of degrees of freedom which are entangled
between both parties?  
Terhal and Hodorecki\cite{terhal001} have recently addressed this question
by introducing the concept of {\it Schmidt number} of a density matrix. 
This number characterizes the minimum Schmidt rank of the pure states 
that are needed to construct such density matrix.
Furthermore, they proved that the Schmidt number is 
non-increasing under local operations and 
classical communication, i.e. 
it provides a legitimate entanglement measure, or more precisely a
monotone \cite{nielsen}. 
 Finally, they introduced also the concept
of $k$-positive maps
which witness the Schmidt number, 
in the same way that positive maps witness entanglement. 
Recently, the concept of 
Schmidt rank and mean Schmidt number has been extended to pure\cite{schmidtnu} 
and mixed states\cite{eisert001}
of multipartite systems.

Let us recall that a map  is called positive (PM) if
it maps positive operators into positive operators. 
A necessary and sufficient criterion for
separability  of a density matrix $\rho$
was introduced
by the Horodeckis \cite{horo96} in terms of PM's.
Their criterion asserts that  a state $\rho$ acting on a composite Hilbert
space ${\cal H}_A\otimes{\cal H}_B$ is separable iff the tensor product of 
any positive map  acting on $A$ and the 
identity acting on $B$ (or  vice versa) maps $\rho$ onto a
positive operator. 
This criterion, however, involves the characterization 
of the set of all PM's, which is  {\it per se} a formidable task.
Similarly, the
characterization of the set of $k$-positive maps\cite{terhal001} is a completely
open problem. A complementary approach to study entanglement,
introduced by
Terhal\cite{terhal002},
 is based on the so-called  entanglement witnesses (EW). An entanglement
witness $W$ is an observable that reveals the entanglement of some entangled
state $\rho$, i.e. $W$  is such that $\rm{Tr}(W\sigma)\ge 0$ for all 
separable $\sigma$, 
but Tr$(W\rho)<0$. The
Hahn-Banach theorem implies that a state $\rho$ is
entangled iff there exists a witness that detects it\cite{horo96}.
There is an isomorphism between positive maps and
entanglement witnesses\cite{Ja72}.  

A well-known example of a
positive map is the transposition $T$: its tensor extension is
the partial transposition (PT)
$I\otimes T$ (see \cite{pt}). 
This map is positive on all separable states\cite{peres96}, and
obviously detects all the entangled states that have non positive
partial transposition (termed NPPT).
However, given a PPT entangled state (PTTES),
i.e. a state with bound entanglement \cite{horo98},
it is in general very difficult to find an EW that detects it. 
A major step in the characterization of both, EW's and the minimal set of 
them which are needed to detect all
entangled states, has been presented in \cite{lewen001}.

In this paper we extend the notion of entanglement witnesses (EW)
to Schmidt number $k$ witnesses ($k$-SW), where $k\geq 2$.     
To this aim we define an observable 
which  is non-negative (negative)
for all (at least one)  $\rho$ of Schmidt number $k-1$ ($k$).
Following \cite{lewen001}, we  express 
such operators in their canonical form, and show how
to optimize them. 
Using this approach we obtain novel insight in the 
structure of the set of PPT-bound entangled states,
determining the minimum number of degrees of freedom that must be 
entangled in order to prepare them.
We present  strong evidence that all PPTES
in $3\times 3$ systems have Schmidt number 2. In $N\times M$ systems 
($N\ge M$) we expect PPTES-states to have a Schmidt number $k<M$ in 
contrast with non-PPT entangled states that can have any Schmidt
number $2\le k\le M$. 
Before going into the details of the paper
we recall the definitions of the Schmidt rank of a pure state $\ket{\psi}$, and
the Schmidt number of a density matrix $\rho$:
\newtheorem{guess}{Definition}
\bg 
{\rm A bipartite pure state $\ket{\psi}\in {\cal H}_A\otimes {\cal H}_B$,
where dim${\cal H}_A=M$  and dim${\cal H}_B= N\ge M$,
has Schmidt rank $r$ if its Schmidt decomposition reads
$\ket{\psi}=\sum_{i=1}^{r}a_i\ket {e_i}\ket {f_i}
\label{schrank} $,
where  $r\le M$ ,
$\sum_i^{r}a_i^2=1$, and $a_i>0$.}
\eg

\bg
{\rm Given the  density matrix $\rho$ of a bipartite system  and all its
possible decompositions in terms of pure states, namely
$\rho=\sum_i p_i \ket{\psi_i^{r_i}}\bra{\psi_i^{r_i}}$,
where $r_i$ denotes the Schmidt rank of $\ket{\psi_i}$, the Schmidt number
of $\rho$, $k$, is defined as $k={\rm{min}}\{r_{\rm{max}}\}$
where ${r}_{\rm{max}}$ is the maximum Schmidt rank within a decomposition,
and the minimum is taken over all decompositions
\cite{terhal001}}.
\eg

Let us denote the whole space of density matrices in $N\times M$ by $S_M$,
and the set of density matrices that have Schmidt number $k$ or less 
by $S_k$. $S_k$ is a convex compact subset of $S_M$\cite{terhal001};  
a state from $S_k$ will be called a state of (Schmidt) class $k$.
Sets of increasing Schmidt number are embedded into each other,
i.e. $S_1\subset S_2\subset...S_k ...\subset S_M$.
In particular, $S_1$ is the set of separable states
(i.e. those that can be written as a convex combination of product states);
$S_2$ is the set of entangled states of Schmidt number 2, 
i.e. those with only two degrees of freedom between the
two parties being entangled, etc.

To determine which is the Schmidt number of a density matrix $\rho$
notice that due to the fact that the sets $S_k$ are convex and compact, 
any arbitrary density matrix of class $k$  can be decomposed as a 
convex combination of a density matrix of class $k-1$ and  a
remainder $\delta$\cite{Le98}:

\newtheorem{guess2}{Proposition}
\begin{guess2}
\rm { Any state of class $k$, $\rho_k$,  can be
written as a convex combination of a density matrix of class $k-1$ and
a so-called $k-$edge state $\delta$:
\be
\rho_k  =(1-p) \rho_{k-1} + p \delta, \;\;\;\, 1\ge p> 0
\label{decom}
\ee
where the edge state $\delta$ has Schmidt number $\ge k$.}
\end{guess2}

The decomposition (\ref{decom}) is obtained by
subtracting projectors onto pure states of Schmidt rank inferior to
$k$, $P=\ket{\psi^{<k}}\bra{\psi^{<k}}$
such that $\rho_k-\lambda P\ge 0$. Here $\ket{\psi^{<k}}$ stands for 
pure states of Schmidt rank $r <k$.
Denoting by $K(\rho)$, $R(\rho)$, and
$r(\rho)$ the kernel, range, and rank of $\rho$ respectively,   
we observe that $\rho'\propto \rho-\lambda {\kepsi}{\brapsi}$  
is non negative iff ${\kepsi}\in R(\rho)$ 
and $\lambda\le {\brapsi}\rho^{-1}{\kepsi}^{-1}$ (see \cite{Le98}).
The idea behind this decomposition is that the edge state $\delta$
which has generically lower rank contains all the information
concerning the Schmidt number $k$ of
the density matrix $\rho_k$. 

Note that there 
exists an optimal decomposition of the form (\ref{decom}) with $p$ minimal.
Also restricting ourselves to decompositions
$\rho_k=\sum_i p_i \ket{\psi_i^{r_i}}\bra{\psi_i^{r_i}}$
with all $r_i\le k$, we can always find a decomposition
of the form (\ref{decom}) with  $\delta\in S_k$. We define below more
precisely what an edge state is.

\bg
{\rm A $k$-edge state $\delta$
is a state such that
$\delta-\epsilon {\kepsi}{\brapsi}$ is not positive,
for any $\epsilon>0$ and $\ket{\psi^{<k}}$.} 
\eg
\newtheorem{guess1}{Criterion}
\begin{guess1} \rm{A mixed state
$\delta$ is a $k$-edge state iff there exists no
 $\kepsi$ such that  $\kepsi\in R(\delta)$.}
\end{guess1}
Let us now define a $k$-Schmidt witness ($k$-SW):  
\bg
\rm{ A hermitian operator $W$ is a Schmidt witness of class $k$ 
iff Tr$(W\sigma)\ge 0$ for all $\sigma\in S_{k-1}$, and there exists 
at least one $\rho \in S_k$ such that Tr$(W\rho)< 0$.}
\eg
Notice that detecting inseparability is, thus, equivalent to
searching witnesses of Schmidt class 2.
Also, the problem of 
distillability \cite{lewen002,horo98,bennet96,durr001} can be recast 
in the language of witnesses of Schmidt number 2 and 3,
{\it i.e.} if $\rho^{T_B}$ is 
a 2-SW (3-SW) then $\rho$ is a distillable (1-copy nondistillable) state.
It is straightforward to see that every SW that detects $\rho$
given by (\ref{decom}) also detects the edge state $\delta$,
since if Tr$(W\rho)<0$  then necessarily Tr$(W\delta)<0$, too. 
Thus, the knowledge of  all SW of $k$-edge states fully characterises 
all $\rho\in S_{k}$. Below, we show how to
construct for any edge state a SW which detects it.
Most of the technical proofs used to construct and optimise
Schmidt witnesses are very similar to those presented in 
Ref.\cite{lewen001} for entanglement witnesses.

Let $\delta$ be a $k$-edge state,
$C$ an arbitrary positive operator such that ${\rm Tr}(\delta
C)>0$, and $P$ a positive operator whose range fulfills
$R(P)=K(\delta)$. We define
$\label{epsilon1} 
\epsilon\equiv  \inf_{\kepsi}{\brapsi}P{\kepsi}$ and 
$ c\equiv \sup {\bra{\psi}}C{\ket{\psi}}$. 
Note that $c>0$ by construction and $\epsilon > 0$, 
because $R(P)= K(\delta)$ and 
therefore, since $R(\delta)$ does not contain any ${\kepsi}$ 
by the definition of edge state, $K(P)$ cannot contain any  
${\kepsi}$ either. This implies:
\newtheorem{guess3}{Lemma}
\begin{guess3}
{\rm
Given an $k$-edge state $\delta$, then
\be
W = P-{\epsilon \over c}C
\ee
is a $k$-SW which detects $\delta$.} 
\end{guess3}
The simplest choice of $P$ and $C$ consists in taking 
projections onto $K(\delta)$ and the identity
operator, respectively. As we will see below, this
choice provides us with a canonical form for a $k$-SW. 

\begin{guess2} 
{\rm
Any Schmidt witness can be written it the {\it canonical} form:
\be
W=\tilde{W}-\epsilon \one\ ,
\ee
such that $R(\tilde W)= K(\delta)$, where $\delta$ is a
$k$-edge state and $0<\epsilon\le {\rm inf}_{|\psi\rangle\in
S_{k-1}}{\bra{\psi}}\tilde{W}{\ket{\psi}}$}.
\end{guess2}
\noindent {\it Proof:} Assume $W$ is an arbitrary $k$-SW so
$W$ has at least one negative
eigenvalue. Construct $W+\epsilon \one =\tilde{W}$, so $\tilde{W}$ is a
positive operator, but it does not have a full rank 
$K({\tilde{W}})\neq\emptyset$ (by continuity this construction 
is always possible). But 
${\brapsi} \tilde{W}{\kepsi} \ge \epsilon >0$ since $W$ is a $k$-SW,
ergo no ${\kepsi}\in K(\tilde{W})$.$\Box$\\
Let us now introduce some additional notations.

\bg
{\rm A $k$-Schmidt witness $W$ is {\it tangent} to $S_{k-1}$ at $\rho$ 
if $\exists$ a state $\rho\in  S_{k-1}$ such that 
Tr$(W\rho)=0$}.
\eg

\newtheorem{guess4}{Observation}
\begin{guess4}

\rm{The state $\rho$ is of Schmidt class $k-1$ iff
for all $k$-SW's tangent to $S_{k-1}$, Tr$(W\rho)\ge 0$.}
\end{guess4}

\noindent{\it Proof} (See \cite{lewen001}): (only if) Suppose that $\rho$ is
of class $k$. From Hahn-Banach theorem, $exists$ a $k$-SW $W$, that
detects it. We can subtract 
$\epsilon \one$ from $W$, making $W-\epsilon \one$ tangent to 
$S_{k-1}$ at some $\sigma$, but then Tr$(\rho(W-\epsilon\one))<0$.$\Box$

We will now discuss the optimisation of a Schmidt witness.
As proposed in \cite{lewen001}(a) an entanglement witness W is optimal 
if there exists no other EW that detects more states than it. 
The same definition can be applied to Schmidt witnesses. 
We say that a $k-$Schmidt witness $W_2$ is  finer   than  
a $k-$Schmidt witness $W_1$, if 
$W_2$ detects more states than $W_1$. Analogously, we define a
$k-$Schmidt witness  $W$ to be optimal when 
there exists no finer witness than itself. 
Let us define the set of ${\kepsi}$ 
for which the expectation value of the
$k$-Schmidt witness $W$ vanishes:
\be
T_{W}=\{ \kepsi \st \brapsi W \kepsi=0 \}\ ,
\ee
i.e. the set of tangent pure states of Schmidt rank $<k$.
$W$ is an optimal $k$-SW iff $W-\epsilon P$ is not a $k$-SW,
for any positive operator $P$. If the set
$T_{W}$ spans the whole Hilbert space, then $W$ is an optimal $k$-SW.
If $T_{W}$ does not span ${\cal H}_A\otimes {\cal H}_B$, 
then we can optimize the witness by subtracting from it a positive operator
$P$, such that $PT_{W}=0$. 
This is possible, provided 
$\inf_{|e_1\rangle,|e_2\rangle 
\in{\cal H}_A}[P_{e_1e_2}^{-1/2}W_{e_1e_2}P_{e_1e_2}^{-1/2}]_{\rm min}>0$, 
where for any $X$ acting on ${\cal H}_A\otimes{\cal H}_B$
\be
X_{e_1e_2}=\left[\begin{array}{cc}\langle e_1|X|e_1\rangle & \langle
e_1|X|e_2\rangle\\ \langle e_2|X|e_1\rangle & \langle
e_2|X|e_2\rangle\end{array}\right], 
\ee
acts in $C^2\otimes{\cal H}_B$, and $[X]_{\rm min}$ denotes its
minimal eigenvalue (see\cite{lewen001}).
An example of an optimal witness of Schmidt number $k$ 
in $H_m\otimes H_m$ is given by
\be
W=\one-\frac{m}{k-1}\calp \;,
\label{example}
\ee
where $\calp$ is a projector onto a maximally entangled state
$|\Psi_+\rangle=\sum_{i=0}^{m-1}\ket{ii}/\sqrt{m}$. 
The $k$-positive map corresponding to (\ref{example})
has been  discussed in \cite{terhal001}. For $k=3$ and $m\ge 3$, 
the partial transpose of (\ref{example}) provides an example
of a one copy non--distillable state with  non-positive partial transpose 
\cite{durr001}. Note that
$W$ is decomposable, i.e. $W=\tilde P+\tilde Q^{T_A}$,
where $\tilde P,\tilde Q\ge 0$, and therefore
it cannot detect any PPTES\cite{lewen001}(a). This can be seen 
by rewriting (\ref{example})
as $W=(1-1/k)\one+2P_a^{T_A}/k$, where $P_a^{T_A}$ is the partially 
transposed projector onto the antisymmetric
subspace of ${\cal H}_m\otimes {\cal H}_m $.

Let us now focus on the case ${\cal H}_3\otimes {\cal H}_3 $ (two qutrits). 
We summarize below the following observations:

\noindent i) Any 2-SW (entanglement witness) has the form $W=Q-\epsilon\one$, 
where $K(Q)$ does not contain any
product vector, i.e. $r(Q)\ge 5$ \cite{kraus001}(b).

\noindent ii) Any 3-SW has the form $W=Q-\epsilon\one$, where $r(Q)=8$. 
This follows from the fact that any
2--dimensional subspace of ${\cal H}_3\otimes {\cal H}_3 $ contains a 
vector of Schmidt rank 2. Note that thus
we have $W=\tilde Q -\epsilon P$, where $P$ is a projector on a vector 
$|\Psi^3\rangle$ of Schmidt rank 3 orthogonal
to $R(Q)$, and $\tilde Q=Q-\epsilon\one_{Q}$ is positive ($\one_Q$  denotes  the 
projector on $R(Q)$).

\noindent iii) Let $A$ be a local transformation in Alice's space
that transforms the maximally entangled state
$|\Psi_+\rangle$ to $|\Psi^3\rangle$, and let the Schmidt coefficients
of $|\Psi^3\rangle$ be $a_1\ge a_2\ge a_3>0$. We can write $W=\tilde Q
 +  (\lambda_{\rm min}-\epsilon)\one-\lambda_{\rm min}AA^{\dag}/3+2\lambda_{\rm min}
(AP_aA^{\dag})^{T_B}/3$, with $\lambda_{\rm min}=[Q]_{\rm min}$.
This implies that if $(\lambda_{\rm min}-\epsilon)\one-\lambda_{\rm
min}AA^{\dag}/3$  is positive definite, i.e.
$\lambda_{\rm min}(1-a_1^2)\ge \epsilon$, then $W$ is decomposable.  On the other hand, 
we observe that for
$|\Psi^2\rangle$ such that 
$|\langle\Psi^2|\Psi^3\rangle|^2=a_1^2+a_2^2$, 
we have
$0\le\langle\Psi^2|W|\Psi^2\rangle\le \lambda_{\rm max}a_3^2-\epsilon$, where 
$\lambda_{\rm max}=[Q]_{\rm
max}$. In turn, these two observations imply:

\begin{guess3}
\rm{
If $\lambda_{\rm max}/\lambda_{\rm min} \le 1+a_2^2/a_3^2$, then $W$ is decomposable.}
\end{guess3}

Note that if $W$ does not fulfill the assumption of this Lemma, it is very
likely that it can be transformed
using local transformations to fulfill it. These observations 
allow us to formulate the following conjecture:

\newtheorem{guess5b}{Conjecture}
\begin{guess5b}
\rm{
In ${\cal H}_3\otimes{\cal H}_3$ all PPT entangled  states 
have Schmidt number 2, i.e. all Schmidt
witnesses of class 3 are decomposable.}
\end{guess5b}
\noindent{\it Evidence:} Obviously, it suffices to prove 
the conjecture for the edge states.
First we prove it rigorously
for rank 4 edge states, such as those constructed from unextendible product
bases\cite{Be98}(a), chessboard states of Ref.\cite{Be98}(b), 
and generalized Choi matrices \cite{Be98}(c).

\begin{guess3}
\rm{
All PPT entangled states of rank 4 have Schmidt number 2.}
\end{guess3}

\noindent{\it Proof:} If $r(\delta)$=4 then there exists a product vector 
$|e_1,f\rangle\in K(\delta)$\cite{kraus001}(b). 
From
$\delta^{T_A}\ge 0$ we see that  $|e^*_1,f\rangle\in K(\delta^{T_A})$. 
Let $|e_i\rangle$, $i=1,2,3$
form an orthonormal basis in ${\cal H}_A$. We have then 
$\langle e_1|\delta|e_i,f\rangle=0$ for $i=2,3$.
Thus, $\delta|e_2,f\rangle=|\Psi^2\rangle=|e_2,g\rangle +|e_3,h\rangle$, i.e. 
$|\Psi^2\rangle$ has Schmidt rank 2. We
can write then $\delta=\delta'+\Lambda|\Psi^2\rangle\langle\Psi^2|$, where 
$\delta'\ge 0$, $\Lambda^{-1}=
\langle \Psi^2|\delta^{-1}|\Psi_2\rangle= \langle \Psi^2|e_2,f\rangle$. Note 
that $r(\delta')=3$, and $\delta'
|e_i,f\rangle=0$ for $i=1,2$, while $\delta'
|e_3,f\rangle=(\delta-\Lambda|\Psi^2\rangle\langle\Psi^2|)|e_3,f\rangle=
|\Phi^2\rangle=|e_2,\tilde g\rangle
+|e_3,\tilde h\rangle$, and $\ket{\Phi^2}$ has at most Schmidt rank 2. 
This allows
us to  write $\delta'=\delta''
+\tilde\Lambda|\Phi^2\rangle\langle\Phi^2|$, where $\delta''\ge 0$, $r(\delta'')=2$, 
and $\delta''|e_i,f\rangle=0$
for $i=1,2,3$. But, that means that $\delta''$ acts in a $3\times 2$ space  
(orthogonal to $|f\rangle$ in 
${\cal H}_B$), {\it ergo} $\delta''$ (and therefore $\delta'$ and $\delta$)
have Schmidt number 2.$\Box$

From \cite{kraus001} we know that the edge states in ${\cal
H}_3\otimes{\cal H}_3$ have ranks   $r(\delta)+r(\delta^{T_A})\le 13$.
Considering pairs $(r(\delta),r(\delta^{T_A}))$, we observe:

\begin{guess3}
{\rm Typically, for any decomposable EW, $W$ tangent to the set 
of PPTES at the edge state $\delta$
with $(r(\delta),r(\delta^{T_A}))=(5,7),(5,8),(6,6),(6,7),(7,6)$, or $(8,5)$, for 
any $\epsilon>0$,
the non--decomposable witness $W_{\epsilon}=W-\epsilon\one$ is  not
a Schmidt witness of
$S_3$, i.e. there exist a vector $|\Psi^2\rangle$ of Schmidt rank 2, 
such that $\langle \Psi^2|W_{\epsilon}|\Psi^2\rangle< 0$.}
\label{lema4}
\end{guess3}
\noindent To prove it, we first write $W=P+Q^{T_A}$,with  $P,Q\ge
0$, where $R(P)=K(\delta)$, $R(Q)=K(\delta^{T_A})$\cite{lewen001}. 
We then consider $|\Psi\rangle=|e_1,f_1\rangle+\beta|e_2,f_2\rangle$, 
such that $P|\Psi\rangle=0$, $Q|e_i,f_i\rangle=0$
for $i=1,2$. Then, $\langle\Psi|W|\Psi\rangle=2{\rm Re}
(\beta\langle e_2^*,f_1|W|e_1^*,f_2\rangle$. Choosing
the phase of $\beta$ appropriately, we can  always get $\langle\Psi|W|\Psi\rangle\le 0$,
 i.e. $W-\epsilon\one$
cannot be a 3-SW. Let us check if such $|\Psi\rangle$ exists. 
The set of $|\Psi\rangle$'s 
forms a 9 dimensional complex
manifold. The vector $|\Psi\rangle$ has to fulfill 
$L=r(P)+2r(Q)=27-r(\delta)-2r(\delta^{T_A})$ equations, and one inequality
for the phase of $\beta$. Obviously, $L<9$ for 
$(r(\delta),r(\delta^{T_A}))=(5,7),(5,8),(6,7)$, and $(7,6)$, 
so that we expect  to have an infinite family of solutions, and in particular 
those with the desired phase of
$\beta$. 
While examples of edge states with ranks $(5,8),(5,7)$ 
are not known, the Horodecki matrix 
of Ref. \cite{horo97}, and the matrix from the $\alpha$--family of states of
Ref. \cite{horo98} with $\alpha=4$
have ranks $(6,7)$. We have checked that for those matrices the
desired $|\Psi\rangle$  exists.   
For
$(r(\delta),r(\delta^{T_A}))=(6,6)$, and 
$(8,5)$,
$L=9$ and we expect a finite number of solutions, 
but still some of them fulfilling the 
requirements for $\beta$. 
We conclude that if a Schmidt witness of the class 3 was
non-decomposable, then it could not be of the form  
$W=P+Q^{T_A}-\epsilon\one$, where $P$  is supported on
$R(\delta)$  and $Q$ on $R(\delta^{T_A})$, for 
$\delta$ of the category considered in Lemma 4. The only possibility is that
$(r(\delta),r(\delta^{T_A}))=(5,5),(5,6),(6,5)$, or $(7,5)$. To investigate
these cases we prove:

\begin{guess4}
{\rm
For any edge state $\delta$ with $r(\delta)+r(\delta^{T_A})\le 13$, there
exists an edge state $\tilde\delta$ with
$r(\tilde\delta)+r(\tilde\delta^{T_A})=13$ arbitrarily close to $\delta$ in the
sense of an operator norm.} \label{obser2} 
\end{guess4}
\noindent{\it Proof:} Let us consider for instance the case $(5,5)$. We can add to 
$\delta$ an
 infinitesimally small separable state composed of 2 projectors on product vectors 
from $R(\delta)$ and 
2 from $R(\delta^{T_A})$, making the resulting state $\rho$ of the category $(7,7)$. 
For such state there
exists a finite number of product vectors  $|e,f\rangle\in R(\delta)$,  
$|e^*,f\rangle\in
R(\delta^{T_A})$. We subtract a projector on one such vector, keeping the 
remainder non--negative and
PPT\cite{kraus001}. 
We choose a vector different from the ones used to construct
$\delta$. Generically, the resulting   state will be arbitrarily close 
to $\delta$, but will
have ranks $(6,7)$,
or $(7,6)$.$\Box$

From Observation \ref{obser2} we immediately get that if $\delta$ with ranks 
$r(\delta)+r(\delta^{T_A})\le 13$ does not belong to $S_2$, 
then there would be a state with ranks $\tilde\delta$
with $r(\tilde\delta)+r(\tilde\delta^{T_A})=13$ arbitrarily close to
$\delta$, which, according to the Hahn--Banach theorem
would  not belong to $S_2$ neither.  
But, that contradicts Lemma \ref{lema4}. 
In effect, if Lemma \ref{lema4} is rigorous, then the conjecture is true.

Summarizing, we have presented a general characterization of  witnesses of
Schmidt number $k$, and the methods of optimizing them. The results allow us
to provide strong evidence  that all bound entangled states
with positive partial transpose in two qutrit systems have Schmidt number 2, i.e. can be prepared using
a two qubit entangled state, local operations and classical communication. 

We thank I. Cirac, P. Horodecki, B. Kraus, M. Nielsen and M. Plenio
for discussions. This work has been supported  by the DFG 
(SFB 407 and Schwerpunkt ``Quanteninformationsverarbeitung"), 
the ESF PESC Programme on Quantum Information, and Benasque Center for 
Science Workshop 2000.


\begin{references}
\bibitem{pt}
Given an operator $X$ and an orthonormal basis
$\{|k\rangle\}_{k=1}^N\in H_B$, one defines the partial
transpose of $X$ with respect to $B$ in that basis as follows:
$
X^{T_B} = \sum_{k,k'=1}^N |k'\rangle_B\langle
k|X|k'\rangle_B\langle k|$.
We say that $\rho$ has a positive partial
transposition (PPT) if $\rho^{T_B}\ge 0$. Note that this
property is basis-independent.
\end{references}
\end{document}